\documentclass[journal]{IEEEtran}

\usepackage[dvips]{graphicx}
\graphicspath{{./}}
\DeclareGraphicsExtensions{.eps}
\usepackage[cmex10]{amsmath}
\usepackage{amsfonts}
\begin{document}

\title{Game theoretic approach for end-to-end resource allocation in multihop cognitive radio networks}



\author{Mar\'ia~Canales,
        Jorge~Ort\' in,~\IEEEmembership{Student~Member,~IEEE,}
        and~Jos\'e~Ram\'on~G\'allego
\thanks{The authors are with the Arag\' on Institute of Engineering Research (I3A), University of Zaragoza, Zaragoza E-50018, Spain (e-mail: \{mcanales,jortin,jrgalleg\}@unizar.es).}
\thanks{This work has been supported by the Spanish Government (FPU grant to the second author and Project TEC2011-23037 from the Ministerio de Ciencia e Innovaci\'on (MICINN) and Fondos Europeos de Desarrollo Regional (FEDER).}}

\maketitle

\begin{abstract}
This paper presents a game theoretic solution for end-to-end channel and power allocation in multihop cognitive radio networks analyzed under the physical interference model. The objective is to find a distributed solution that maximizes the number of flows that can be established in the network. The problem is addressed through three different games: a local flow game which uses complete information about the links of the flow, a potential flow game requiring global network knowledge and a cooperative link game based on partial information regarding the links of the flow. Results show that the proposed link game highly decreases the complexity of the channel and power allocation problem in terms of computational load, reducing the information shared between the links forming each flow with a performance similar to that of the more complex flow games.
\end{abstract}


\section{Introduction}
\IEEEPARstart{A}{utonomous}, self-configuring multihop networks present a versatile solution to provide broadband services with infrastructure-less deployments and decentralized management. Furthermore, their intrinsic adaptability and resilience can be enhanced with cognitive radio technology \cite{Mitola99}, enabling the nodes of the network to adjust their transmitting parameters to the specific operational environment of the network. One of the main research challenges in these kind of networks is the proposal of efficient and distributed radio resource management solutions that accomplish the channel and power allocation for the links of each flow in the network. 

These solutions should be simple enough to be implemented in real systems, where overall information on the environment is not assured, and should get good results in terms of the global network performance. To tackle these challenges, game theory has recently received an increasing interest in the context of cognitive networks \cite{Wang10}. Game theory is a mathematical tool that analyzes the strategic interactions among multiple decision makers. This characterization of the problem facilitates the evaluation of the expected performance of the network, giving an insight into its behavior. The joint channel and power allocation problem for cognitive radio networks have been already studied with a game theoretical perspective \cite{Nie06}\nocite{Li07}\nocite{Chen08}-\cite{Canales11}. These works focus on the establishment of single hop links between pairs of nodes as isolated entities. However, in a multihop network, the successful establishment of an end-to-end flow between two nodes requires the joint activation of all the links of the flow. In \cite{Gao09} channel allocation for multi radio wireless networks is studied from a game-theoretical perspective, taking into account these end-to-end requirements. However, the problem is analyzed under a simplified propagation model with fixed transmission and sensing ranges, assuming that all flows reside in a single collision domain. Aspects such as the transmission power or the quality of the links, measured through the received Signal-to-Interference-and-Noise-Ratio (\textit{SINR}) are not taken into account.

The aim of this work is to model the behavior of competing flows within an autonomous multihop wireless network using game theory in order to provide strategies that maximize the number of flows that can be established. To the authors' knowledge, this is the first game theoretic approach to model the end-to-end resource allocation problem in multihop cognitive radio networks under a realistic physical interference model \cite{Gupta00}. Specifically, we propose, define and evaluate with simulations three different games that perform an end-to-end joint channel and power allocation in this scenario. In two of them, the flows themselves are the players, which implies high communication and computation complexities. In the third one, the individual links of each flow are the players, so each link updates its strategy locally, with a certain degree of cooperation among links, but with much lower complexity than the flow games. The obtained results demonstrate that this simple cooperative link game can provide a feasible solution for the addressed problem with good performance, limited computational complexity and low requirements of environmental information. This work is organized as follows. In Section II the system model is described. Section III presents the game theoretic approach with the description of the proposed games. The simulation framework and the obtained results are shown in Section IV. Finally, in Section V the main conclusions are summarized.

\section{System Model}

The system considered in this work is a multihop wireless network with $N$ nodes. The maximum transmission power for each node is $P_{\text{\textit{max}}}$. This transmission power is discretized into a finite number of $Q$ levels, equispaced between 0 and $P_{\text{\textit{max}}}$. There are $C$ non-interfering channels in the network, and each node can only transmit in a specific subset of these channels, which can vary from one node to another. Within this network, a set $F$ of flows is desired to be established. Each flow $f$ is formed by a sequence of $k$ directed links between the source node and the destination node obtained with the Dijkstra algorithm. Given a directional link $l$ between a pair of nodes $\left(l_\text{\textit{TX}} \rightarrow l_\text{\textit{RX}}\right)$, the channel gain from transmitter $\left(l_\text{\textit{TX}}\right)$ to receiver $\left(l_\text{\textit{RX}}\right)$ is defined as $g_{l,l} = d_{l,l}^{-\gamma}$, being $d_{l,l}$ the distance from $l_\text{\textit{TX}}$ to $l_\text{\textit{RX}}$ and $\gamma$ the path loss index. Similarly, $g_{m,l} = d_{m,l}^{-\gamma}$ represents the channel gain from the transmitter of link $m$ $\left(m_\text{\textit{TX}}\right)$ to the receiver of link $l$ $\left(l_\text{\textit{RX}}\right)$.

A flow is feasible if and only if all its links can be established. To determine if a link is active or not, the physical interference model has been employed in the work. Under this model, a directional link $l$ can be successfully established if and only if the \textit{SINR} at the receiver $\left(l_\text{\textit{RX}}\right)$ is higher than a certain threshold $\alpha$:

\begin{equation}
\text{\textit{SINR}}_l = \frac{p_l\cdot g_{l,l}}{P_N+\displaystyle\sum_{\substack{m \in L, m \neq l \\ c_m = c_l}}{\!\!p_m\cdot g_{m,l}}} \geq \alpha
\end{equation}

\noindent with $p_l$ the power assigned to link $l$, $c_l$ the channel used by link $l$, $L$ the set of links in the network and $P_N$ the background noise power.

\section{Game Theoretic Approach}

Let be the game $\Gamma = \left\lbrace M,{\left\lbrace S_i \right\rbrace}_{i \in M}, {\left\lbrace u_i \right\rbrace}_{i \in M} \right\rbrace$, where $M$ is the finite set of players, $S_i$ is the set of strategies related to player $i$ and $u_i : S \rightarrow \mathbb{R}$ is the utility function of that player, with $S = { \times _{i \in N}}{S_i}$ the strategy space of the game. This utility function $u_i$ is a function of $s_i$, the strategy selected by player $i$, and of $s_{-i}$, the current strategy profile of the rest of the players. Players will selfishly choose the actions that improve their utility functions considering the current strategies of the other players. One general key issue when designing a game is the choice of $u_i$ so that the individual actions of the players provide a good overall performance. Two properties are usually desirable: the game should have an equilibrium point and this point should maximize the network utility. This equilibrium point, where no player has anything to gain by unilaterally deviating, is known as Nash Equilibrium (NE). Thus, a Nash equilibrium of a game $\Gamma$ is a profile ${s^*} \in S$ of actions such that for every player $i \in M$ we have: 
	 	
\begin{equation}
	 	{u_i}(s_i^*,s_{ - i}^*) \ge {u_i}(s_i^{},s_{ - i}^*)
\end{equation}

\noindent for all ${s_i} \in S$, where $s_i$ denotes any strategy of player $i$ and $s_{ - i}^* \in S$  denotes the strategies of all the players other than player $i$ in the profile ${s^*}$.

Another specific key issue for the problem addressed in this work is the definition of the players of the game. In this regard, we consider two different choices: in the first one, the players of the game are the flows themselves (flow game), while in the second one the players are the set of links belonging to the flows (link game). It must be noted that in the flow game, there must be a physical entity (e. g. the source node of the flow) which actually acts as the player. This entity gathers the set of available channels and powers of each link in the flow and performs the specific channel and power allocation for all of them. Considering the different alternatives of players and utility functions, we propose the following three games to analyze the system:

\subsection{Flow games}

In these games, the players are the $|F|$ flows of the network $(M = F)$. The set of strategies $S_i$ of a flow $i$ with $k$ links is formed by the Cartesian product of the set of strategies $A_l$ of each link of the flow: $S_i = A_1 \times A_2 \times \ldots \times A_k$. The subindexes represent the position of the link in the flow, from the source to the destination. Each link strategy $a_l  = \left(p_l, c_l\right)$ is the allocation of transmission power and frequency channel for that link. 

\subsubsection{Local flow game (LFG)}

The utility function of flow $i$ is directly related to the success in the establishment of the flow. If $L_i$ denotes the subset of links belonging to flow $i$, the utility function is defined as:

\begin{equation}
u_i \left(s_i,s_{-i}\right) = \! \begin{cases}
1 & \! \text{if $\text{\textit{SINR}}_l > \alpha$ $\forall l \in L_i$}\\
-1 & \! \text{if $\exists l \in L_i | \left(p_l>0\right) \wedge \left(\text{\textit{SINR}}_l < \alpha\right)$}\\
0 & \! \text{otherwise (if $p_l = 0$ $\forall l \in L_i$)}
\label{utility}
\end{cases}
\end{equation}

The $-1$ value for any $\text{\textit{SINR}}_l < \alpha$ tries to introduce a degree of cooperation to compensate the inherent selfishness of this game: if a link of the flow cannot be established, the flow is not viable and it is better to stop the transmission of the links in the flow to reduce the interference on the remaining flows. 

In the particular case where all the flows are single-hop routes (i.e., single links), this game can be seen as a local link game (LLG). In this kind of games the existence and convergence to a pure NE cannot be assured \cite{Canales11}, \cite{Andrews09}. Therefore, the same applies for this local flow game, which is an extension of the previous one.

It must be noted that a full knowledge of all the strategy spaces of each link of the flow $(A_l)$ is required in this game to perform the strategy selection. In addition, to obtain the \textit{SINR} of each link, the node of the flow which acts as the player needs to know the channel gain $g_{l,l'}$ from the transmitter of each link $l$ $\left(l_\text{\textit{TX}}\right)$ of the flow to the receivers of all the links $l'$ $\left(l'_\text{\textit{RX}}\right)$ in the flow and the interference levels at all these receivers.

\subsubsection{Potential flow game (PFG)}

An Exact Potential Game is a game for which there exists a potential function $V : S \rightarrow \mathbb{R}$ such that:

\begin{equation}
\begin{split}
\Delta u_i & = u_i\left( s_i,s_{-i} \right) - u_i\left( s'_i,s_{-i} \right) = \Delta V = \\
& =  V \left( s_i,s_{-i} \right) - V \left( s'_i,s_{-i} \right), \; \; \; \forall i \in M,\forall {s_i},s{'_i} \in {S_i}
\end{split}
\end{equation}

If only one player acts at each time step (repeated sequential game) and the acting player maximizes (best response strategy) or at least improves (better response strategy) \cite{Wang10} its utility, given the most recent action of the other players, then the process will always converge to a NE. In addition, global maximizers of the potential function $V$ are NE, although they may be just a subset of all NE of the game. 

We can define a potential game with $V$ equals to the objective to maximize, in this case, the number of flows to be established. A direct option is to define the utility function $u_i$ equal to the potential function (identical interest games \cite{Song08}):

\begin{equation}
u_i\left(s_i,s_{-i}\right) = \displaystyle\sum_{j \in F} \lambda_j
\end{equation}

\noindent where $\lambda_j$ is $1$ if flow $j$ is active and $0$ otherwise.

In this game, besides requiring the $A_l$ for all the links in the flow, each player needs global information about all the flows (and consequently, all the links) in the network. To compute the viability of all the flows for each selected strategy, each player requires the channel gains $g_{l,m}$ between any pair of transmitting and receiving nodes of the links in the network and the current strategies $a_l = \left(p_l, c_l\right)$ of all these links. 

\subsection{Cooperative link game (CLG)}

The main drawback of the flow games is the complexity that they pose in terms of the necessity of sharing information amongst the nodes of the network and the computational load required to perform the strategy selection. The cardinality of the strategy space of each flow is $O((Q\cdot C)^k)$, being the time complexity to select the strategy profile exponential. To solve this problem, we propose a cooperative link game which does not require a central entity in each flow to perform the strategy selection. This decreases both the amount of information shared between the nodes of the network and the complexity of the selection of the strategy profile. Nevertheless a certain degree of cooperation between the links of the flow is required as described below. 

In this game, the players are the set of links belonging to the $|F|$ flows of the network $(M = L)$. The set of strategies $S_i$ of a link $i$ is its available set of power transmission and channel frequency combinations, $s_i = a_i = (p_i, c_i)$. The links belonging to the same flow cooperate in the selection of their strategies. The game is played sequentially from the source to the destination for each link $i$ of a flow $f$, being the strategy selection process divided in two steps. In the first one, each link $i$ selects its strategy according to the following utility function:

\begin{equation}
u_i \left(s_i,s_{-i}\right) = \begin{cases}
1 & \text{if $\text{\textit{SINR}}_l > \alpha$ $\forall l \leq i$}\\
0 & \text{if $p_i = 0$}\\
-1 & \text{otherwise}
\end{cases}
\end{equation}

\noindent i.e., each link tries to choose a strategy that activates its own link without disrupting the previous ones to obtain $u_i = 1$. If this is not possible, it is better for itself and the following links stop transmitting to obtain $u_i = 0$. To compute the \textit{SINR} of the previous links in the flow, each link needs to know the strategies selected by those links (but not the whole strategy space as in the flow games), their channel gains $g_{l',l'}$, the channel gain from the transmitter of the player $i$ to the receivers of those previous links, $g_{i,l'}$ and the interference levels at these receivers. 

In the second step, once all the links of the flow have played, they update their strategy with the following utility function:

\begin{equation}
u_i \left(s_i,s_{-i}\right) = \! \begin{cases}
1 & \!\!\! \text{if $\text{\textit{SINR}}_l > \alpha$ $\forall l \in L_f$}\\
0 & \!\!\! \text{if $p_i = 0$}\\
-1 & \!\!\! \text{otherwise}\\
\end{cases}
\end{equation}

Thus, if all the links have been activated, the flow $f$ is active and no more actions are required. On the contrary, if any link is not active, the flow cannot be established and $u_i = -1$ for all the links. Therefore, to improve their utility and reduce the interference on the remaining flows, all the links in the flow must stop transmitting ($u_i = 0$). 

\subsection{Timing and decision rules}

A repeated sequential game with a round robin scheduling and a better response strategy is considered for all the proposed games. For the flow games (both local and potential), the game is played until a pure NE is found or until a predefined maximum number of iterations is reached. This playing rule requires searching through the entire strategy space, which as stated above implies a high complexity for the flow games. The cooperative link game reduces this complexity at the expense of not evaluating all the possible strategy profiles. Therefore, the profiles corresponding to the NEs may not be reached and consequently a different less restrictive rule is defined to stop the game: if no link has improved its utility (and hence no new flow has been established) in a round robin cycle, the game is stopped. It is worth mentioning that this stopping rule does not imply that some new flow could actually be established since the complete strategy space is not evaluated.

\begin{figure}[!t]
\centering
\includegraphics[width=3.4in, clip=true]{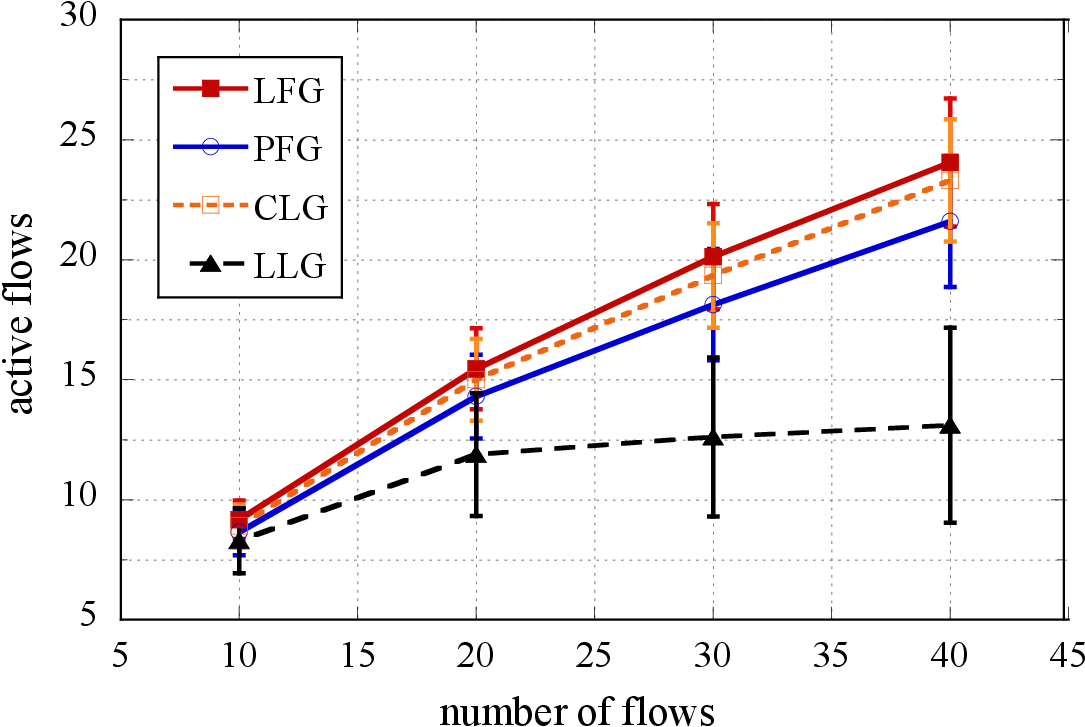}
\caption{Active flows as a function of the number of competing flows in the network for the analyzed games. Mean value and standard deviation (bars).}
\label{fig:active_links}
\end{figure}

\begin{figure}[!t]
\centering
\includegraphics[width=3.4in, clip=true]{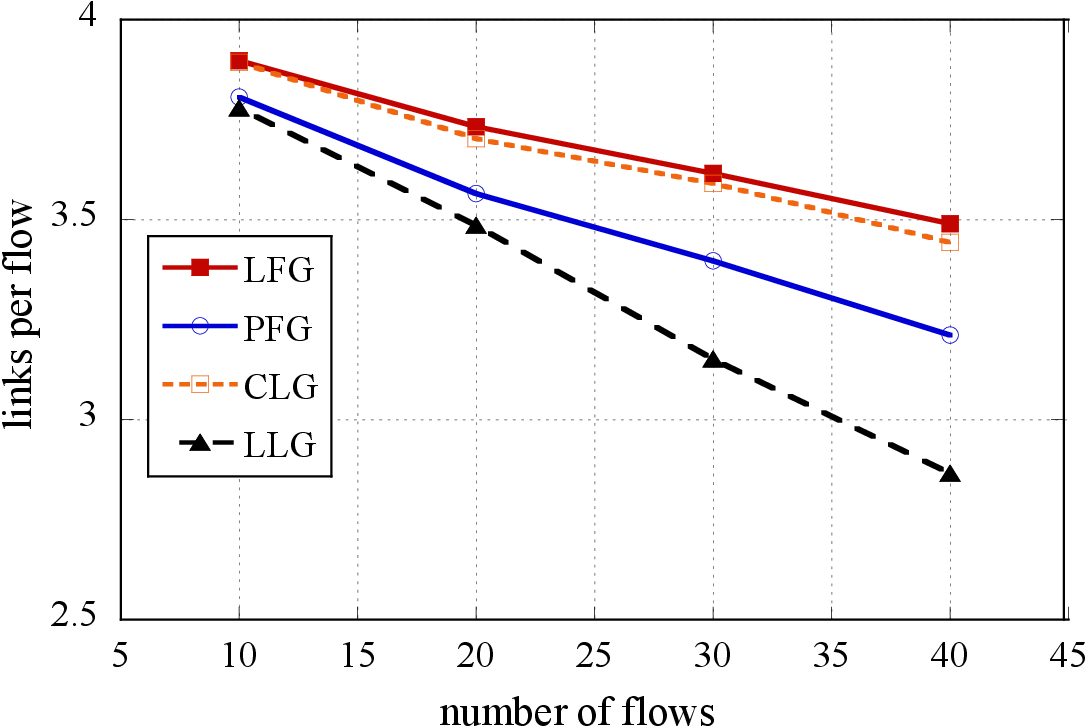}
\caption{Mean number of links per active flow as a function of the number of competing flows in the network for the analyzed games.}
\label{fig:links_in_flow}
\end{figure}

\section{Simulation Results}

The proposed games have been evaluated by simulation. The network consists of 200 nodes deployed in a square area of length 1 km with different random locations. Several numbers of flows ranging from 10 to 40 are generated between nodes of the topology. The source and destination nodes of the flows are also randomly generated. All the presented results are averaged with 100 random instances of the scenario. The maximum number of hops of each flow is limited to 6. The number of non-interfering channels is set to $C = 10$, nevertheless, only a subset of these channels is available to each node. To do so, the scenario is divided into regions of $100\times 100$ m, being assigned only a random subset (between 3 and 8) of these $C$ channels to the nodes in the area. $P_\text{\textit{max}}$ is set to 20 dBm, transmission power is quantized with $Q = 16$ levels and the path loss index is $\gamma = 4$. The \textit{SINR} threshold $\alpha$ is set to 10dB. $P_N$ is set to -70 dBm, which ensures a \textit{SINR} of 10~dB at 100 m. This sets the maximum transmission range in 100 m.

Fig. \ref{fig:active_links} shows the number of active flows obtained with the three proposed games at the equilibrium point (NE for the flow games and the criterion defined in section III.C for the cooperative link game). As a reference, the results obtained with a local link game (LLG) \cite{Canales11} where each link only tries to be established without any knowledge about the flow it belongs to are also included. The three proposed games clearly outperform the LLG, since the only objective of this one is to establish isolated links. In addition, the cooperative link game provides a performance close to that of the local flow game and even better than the potential flow game with a much lower complexity. It must be noted that Fig.~\ref{fig:active_links} only shows the active flows when the game has converged. While the convergence is guaranteed for the potential flow game, it cannot be assured for the local flow game and the cooperative link game. Nevertheless, between 95\% and 98\% of the simulated scenarios for the local flow game and 100\% for the cooperative link game have reached the equilibrium point in the presented results.

Fig. \ref{fig:links_in_flow} shows the mean number of links of the established flows. This number decreases as the congestion in the network (i. e., the number of competing flows) grows. In this situation, the flows with fewer links tend to be established since the requirement that all the links are active in the flow is more easily achieved with short flows.

As for the computational complexity of the games, two variables must be taken into account: the number of steps required to achieve the convergence and the computational load of each step. Table \ref{table:steps} shows the mean number of normalized flow steps for the four games. For the flow games, the flow step is directly the game step, whereas for the link games, the flow step is equivalent to $k$ game steps, being $k$ the number of links (players) of the flow. Therefore, the flow steps for the link games in Table \ref{table:steps} are normalized by dividing the number of link steps by the mean number of links per flow. 

Regarding the computational load of each step, the flow games may require in the worst case exploring $(Q\cdot C)^k$ combinations of channel and power allocations to select the strategy. On the contrary, the worst case for the link game only requires exploring $Q \cdot C$ combinations per link step, that is $k \cdot Q \cdot C$ combinations considering the $k$ links of the flow (the equivalent to a flow game step). Consequently, although the number of flow steps are within the same magnitude order for both flow and link games (Table~\ref{table:steps}), the computational load of each step is much lower in the latter, decreasing the complexity of the game.

\begin{table}[!t]
\renewcommand{\arraystretch}{1.3}
\centering
\caption{Mean and standard deviation (in brackets) of the normalized number of flow steps performed until convergence for the analyzed games.}
\begin{tabular}{|c||c|c|c|c|}
\hline
Flows &  LLG & CLG & LFG & PFG \\
\hline
10 & \centering 44 (11) & 45 (18) & 46 (19) & 30 (4) \\
\hline
20 & 106 (19) & 142 (76) & 171 (94) & 71 (14) \\
\hline
30 & 186 (37) & 286 (140) & 426 (662) & 118 (25) \\
\hline
40 & 271 (52) & 359 (140) & 588 (474) & 166 (32) \\
\hline
\end{tabular}
\label{table:steps}
\end{table}

\section{Conclusion}
In this letter, a game theoretic framework for end-to-end joint channel and power allocation in multihop cognitive radio networks has been proposed and evaluated under the physical interference model. Simulation results have shown that the proposed cooperative link game can provide stable configurations with a global performance similar to more complex flow games. Our future work is focused on a deeper analytical convergence study of the cooperative link game and the definition of new utility functions aimed to reduce fairness problems among short and long flows.



\ifCLASSOPTIONcaptionsoff
  \newpage
\fi


\bibliographystyle{IEEEtran}
\bibliography{./IEEEabrv,./Gallego_CL2011-2574_bibliography}
\end{document}